\documentclass[12pt]{iopart}

\usepackage{graphics}
\begin{document}
\title [Comprehensive Analysis of Neutrinos in SK part II]
{
The Comprehensive Analysis of Neutrino Events Occurring inside the
Detector in the Super-Kamiokande Experiment from the View Point of 
the Numerical Computer Experiments: Part~2\\[20pt]
{\large
   -- L/E Analysis for Single Ring Muon Events I --
}
}
\author{
E. Konishi$^1$, Y. Minorikawa$^2$, V.I. Galkin$^3$, M. Ishiwata$^4$,\\
I. Nakamura$^4$, N. Takahashi$^1$, M. Kato$^5$ and A. Misaki$^6$
}

\address{
$^1$ Graduate School of Science and Technology, Hirosaki University, Hirosaki, 036-8561, Japan }    
\address{
$^2$ Department of Science, School of Science and Engineering, Kinki University, Higashi-Osaka, 577-8502, Japan }
\address{
$^3$ Department of Physics, Moscow State University, Moscow, 119992, Russia}
\address{
$^4$ Department of Physics, Saitama University, Saitama, 338-8570, Japan}
\address{
$^5$ Kyowa Interface Science Co.,Ltd., Saitama, 351-0033, Japan }
\address{
$^6$ Research Institute for Science and Engineering, Waseda University, Tokyo, 169-0092, Japan }
\ead{konish@si.hirosaki-u.ac.jp}

\begin{abstract} 
By referring to the procedures developed in the preceeding paper,
we re-analyze the $L_{\nu}/E_{\nu}$ distribution for 
{\it Fully Contained Events} resulting from quasi-elatic scattering 
(QEL) obtained from the Super-Kamiokande Experiment in relation to 
their assumption that the direction of the incident neutrino coincide 
with that of the produced leptons. 
 As the result of it, we clarify that they do not measure 
$L_{\nu}/E_{\nu}$  distribution itself, but 
$L_{\mu}/E_{\nu}$ distribution 
which cannot show the maximum oscillation existed in the original  
$L_{\nu}/E_{\nu}$ distribution, because 
$L_{\nu}$ could not be approximated by $L_{\mu}$ due to 
the backscattering effect and the azimuthal angle effect in QEL. 

\end{abstract}


\maketitle

\section{Introduction}
In analyzing the observed results in neutrino events occurring inside 
the detector in the Super-Kamiokande Experiment, the direction of the
 incident neutrino is assumed to be coincided with that of the 
produced lepton detected by the detector which is fundamental for 
their analysis.
In order to avoid any misunderstanding toward 
their approximation on the direction of the incident neutrino
we reproduce this approximation 
from the original papers and their related papers in italic.
Kajita and Totsuka state:
\begin{quote}
"{\it However, the direction of the neutrino must be estimated from the
reconstructed direction of the products of the neutrino interaction.
 In water Cheren-kov detectors, the direction of an observed lepton is
assumed to be the direction of the neutrino. Fig.11 and 
Fig.12 show the estimated correlation angle between 
neutrinos and leptons as a function of lepton momentum.
 At energies below 400~MeV/c, the lepton direction has little 
correlation with the neutrino direction. The correlation angle 
becomes smaller with increasing lepton momentum. Therefore, 
the zenith angle dependence of the flux as a consequence of 
neutrino oscillation is largely washed out below~400 MeV/c lepton
momentum. With increasing momentum, the effect can be seen more 
clearly.}" \cite{Kajita1}
\end{quote}
Also, Ishitsuka states in his Ph.D thesis which
 is exclusively devoted into the L/E analysis of the 
atmospheric neutrino from Super Kamiokande Experiment as follows:
\begin{quote}
" {\it 8.4  Reconstruction of $L_\nu$
\vskip 2mm

Flight length of neutrino is determined from the neutrino incident
zenith angle, although the energy and the flavor are also involved.
 First, the direction of neutrino is estimated for each sample by 
a different way. Then, the neutrino flight lenght is 
calclulated from the zenith angle of the reconstructed direction.
\\
\\
 8.4.1 Reconstruction of Neutrino Direction

\vspace{-2mm}

{\flushleft{\underline 
{FC Single-ring Sample}
}
}

\vspace{2mm}

The direction of neutrino for FC single-ring sample is 
simply assumed to be the same as the reconstructed direction of muon.
Zenith angle of neutrino is reconstructed as follows:
\[
\hspace{2cm}\cos\Theta^{rec}_{\nu}=\cos\Theta_{\mu} \hspace{2cm}(8.17) 
\]
,where $\cos\Theta^{rec}_{\nu}$ and $\cos\Theta_{\mu}$ are 
cosine of the reconstructed zenith angle of muon and neutrino,
respectively.}" \cite{Ishitsuka}
\end{quote}
Furthermore, Jung, Kajita {\it et al.} state:
\begin{quote}
"{\it At neutrino energies of more than a few hundred MeV, the 
direction of the reconstructed lepton approximately represents 
the direction of the original neutrino. 
Hence, for detectors near direction of the lepton. Any effects, 
such as neutrino oscillations, that are a function of the neutrino
 flight distance will be manifest in the 
lepton zenith angle distributions.}" \cite{Jung}
\end{quote}
In the present paper, we examine the results on the 
$L/E$ distribution obtained from the 
Super-Kamiokande Experiment in relation to their validity of the 
assumption that the direction of the incident neutrino coincide 
with that of the produced muon. In this case, 
this assumption is expressed as, $L_{\nu}\approx L_{\mu}$. 
Throughout this paper we call $L_{\nu}\approx L_{\mu}$
 as {\it the SK assumption on the direction}.

 In the preceding paper\cite{part1}, we showed that {\it the SK 
assumption on the direction} does not hold even if statistically.
Thus, this conclusion influences straightforwardly on the uncertainty in 
the quantity of $L$ in their $L/E$ analysis. In the present paper,
 we examine the validity of the evidence for the oscillatory signature
 in the atmospheric neutrino oscillation claimed by 
the Super-Kamiokande Collaboration\cite{Ashie1}.
 Super-Kamiokande Collaboration utilize $L_{\mu}$, instead of $L_{\nu}$
 in their $L/E$ analysis, surmising that 
{\it the SK assumption on the direction} may not cause 
the serious error. In order to give a clear answer to "the evidence" 
for the oscillatory nature based on 
{\it the SK assumption on the direction},
in the present paper, we restrict our interest to the analysis of 
single ring muon events due to the quasi-elastic scattering (QEL) among 
{\it Fully Contained Events}, 
namely,

\begin{eqnarray}
         \nu_{\mu} + n \longrightarrow p + \mu^- \nonumber\\
         \bar{\nu}_{\mu}+ p \longrightarrow n + \mu^+, 
\label{eqn:1}
\end{eqnarray}

\noindent because neutrino events which belong this category are of 
experimentally high qualities compared with any other types of the 
neutrino events in their analysis and could give the most clear cut 
conclusions on the neutrino oscillation, if any exist.       
\begin{figure}
\begin{center}
\vspace{-2cm}
\hspace*{-2cm}
\rotatebox{90}{%
\resizebox{0.8\textwidth}{!}{%
  \includegraphics{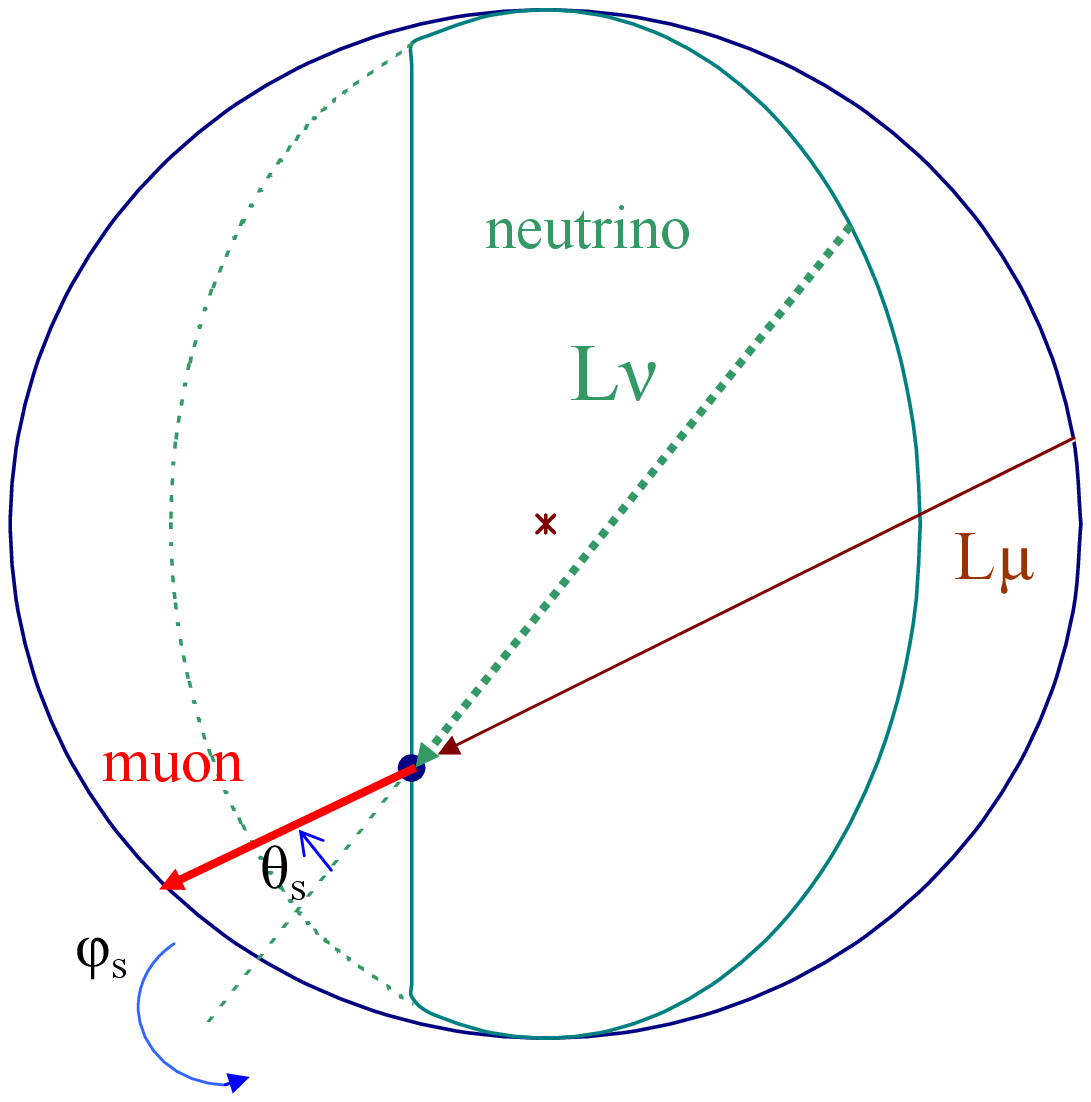}
}}
\vspace{-4cm}
\caption{Schematic view of relations among
$L_{\nu}$, $L_{\mu}$, $\theta_s$ and $\phi_s$
.}
\label{fig:01}       
\resizebox{0.5\textwidth}{!}{%
  \includegraphics{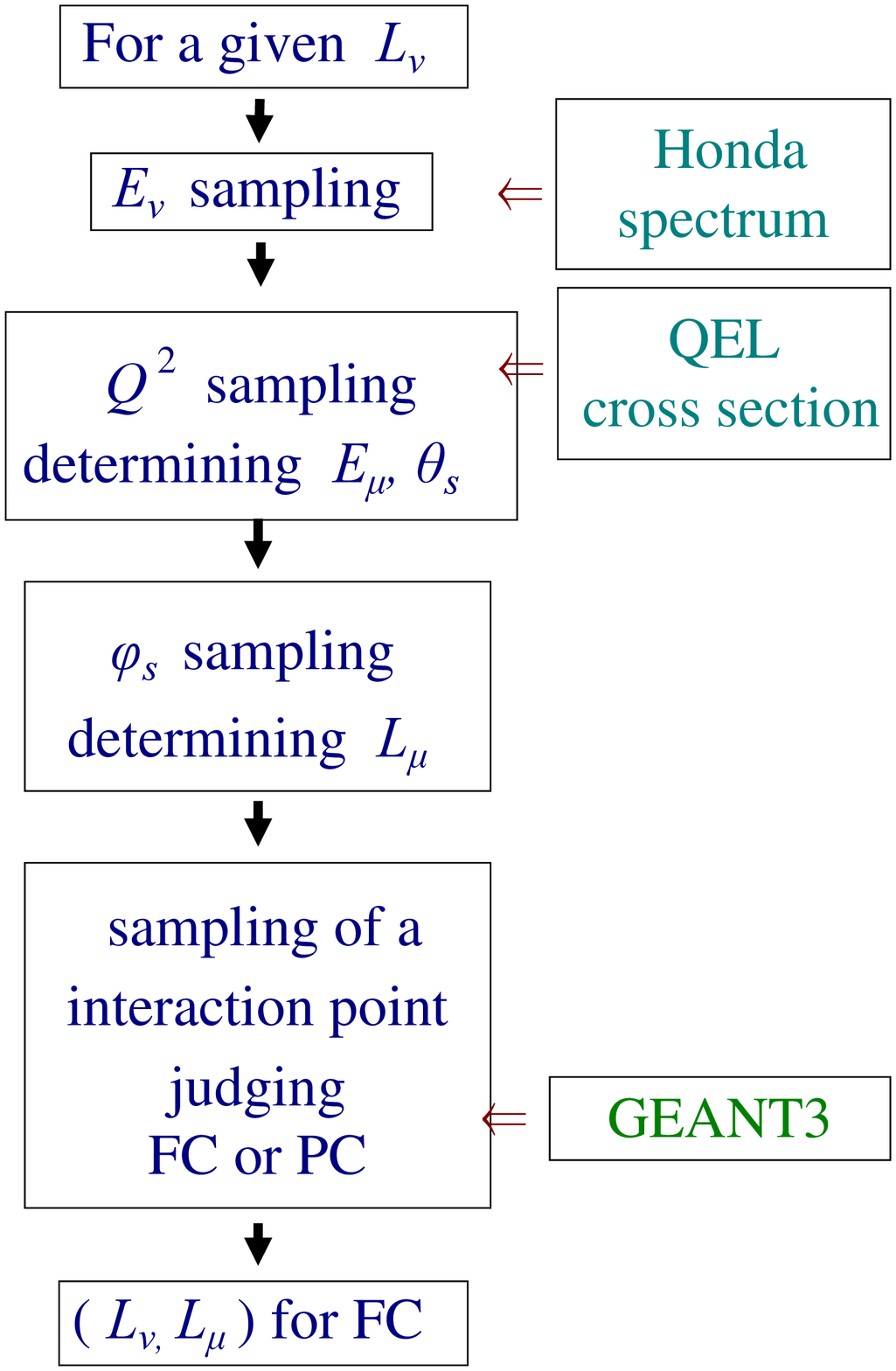}
}
\vspace{-1.5cm}
\caption{The procedure for our numerical experiment
for obtaining $L_{\mu}$ from a given $L_{\nu}$.
}
\label{fig:02}
\end{center}
\end{figure}

In section~2 we give the procedure how to obtain $L_{\mu}$ from a 
neutrino event with given $L_{\nu}$ in stochastic manner.
 In section~3 we give the correlations between $L_{\nu}$ and $L_{\mu}$, 
taking into account of the effect of the backscattering as well as 
the effect of the azimuthal angle on the QEL in stochastic manner.
 As the result of it, we show that 
$L_{\nu}\approx L_{\mu}$,
 {\it the SK assumption on the direction}, 
does not hold even if statistically in both the absence and the 
presence of neutrino oscillation (Figure~3 and Figure~4).
 Also, we treat the correlation between $E_{\nu}$ and $E_{\mu}$, 
in stochastic manner. 
We show that the approximation of $E_{\nu}$ with $E_{\mu}$ 
by the Super-Kamiokande Collaboration does not make so serious 
error compared with the case of $L_{\nu}$ by $L_{\mu}$,
 although their treatment is theoretically unsuitable (Figure~5).

In section~4, we examine 
$L_{\nu}/E_{\nu}$ distribution and show the existence of the maximum 
oscillation under the neutrino oscillation parameters obtained
by the Super-Kamiokande Collaboration, as it must be. 
This fact denotes that our numerical computer experiment is done in correct manner. 

\section{Generation of the Neutrino Events 
Occurring inside the Detector}

In our numerical computer experiment, we obtain 
{\it the Fully Contained Events} resulting 
from QEL in the virtual SK detector,
 the details of which are described in Appendix A.
For the neutrino event with a definite neutrino energy thus generated,
we simulate its interaction point inside the detector and 
the emitted energy of the muon concerned 
which gives its scattering angle uniquely. 
The determination of the neutrino energy, the emitted energy of the muon 
and its scattering angle are described in the preceding paper
 (\cite{part1} see, the Appendices A,B and C in the preceding paper).
  The muon thus generated is pursued in the stochastic manner by using 
GEANT 3 and finally we judge whether the muon concerned stops inside the 
detector ({\it the Fully Contained Event}) or passes through 
the detector ({\it the Partially Contained Event}). 
 For {\it Fully Contained Events} thus obtained, we know the directions 
of the incident neutrinos, the generation points and termination points of 
the events inside the detector, the emitted muon energies,
their scattering angles and 
their azimuthal angles in QEL which lead 
finally their zenith angles, $L_{\nu}$ and $L_{\mu}$
\footnote{
The azimuthal angle, here, is that in QEL, not to that to the earth}.

\begin{figure}
\begin{center}
\vspace{-1cm}
\rotatebox{90}{%
\resizebox{0.6\textwidth}{!}{%
  \includegraphics{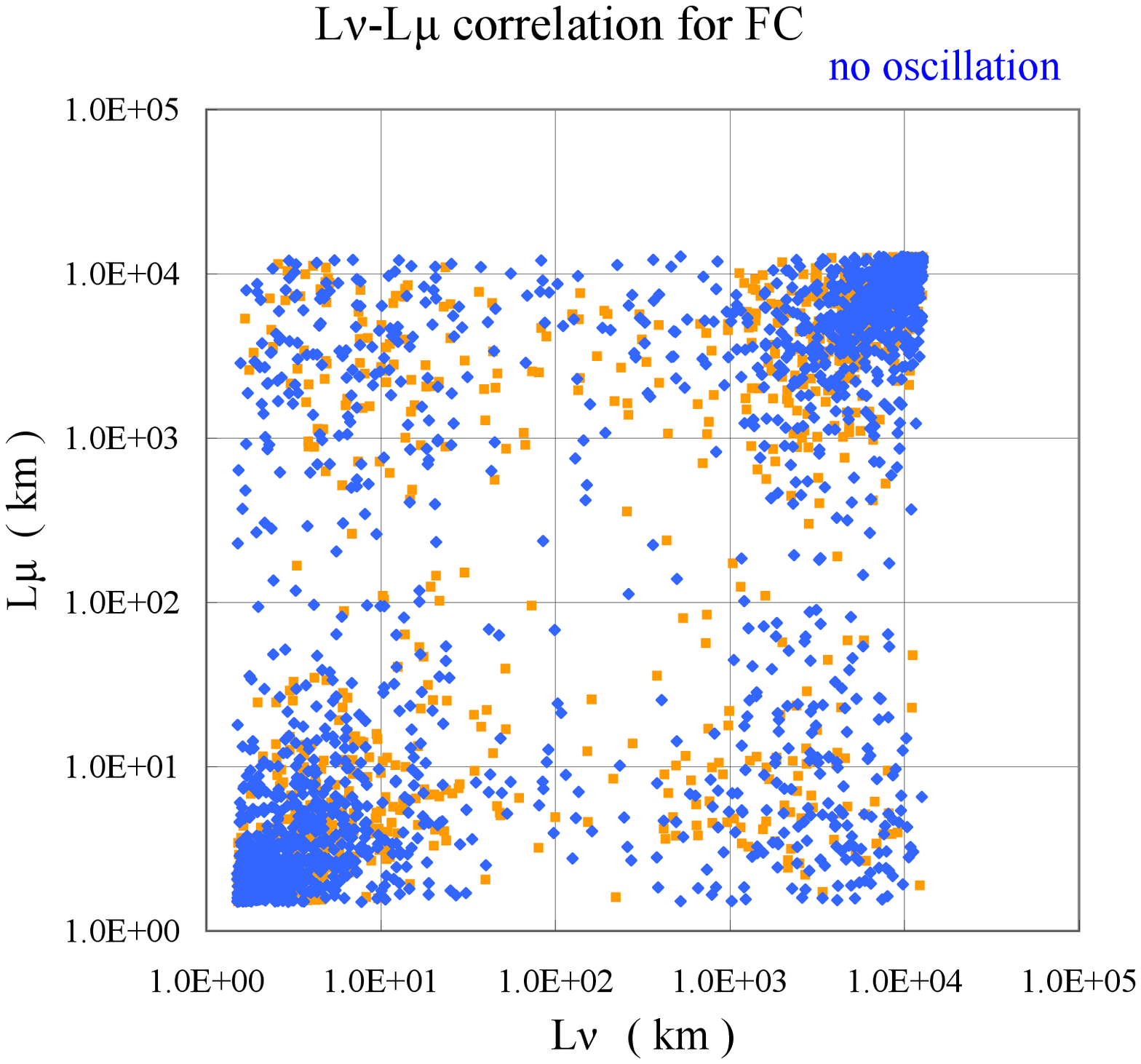}
}}
\vspace{-1.2cm}
\caption{Correlation diagram for $L_{\nu}$ and $L_{\mu}$ without 
oscillation for for 1489.2 live days.
The blue points and orange points denote neutrino events and 
ani-neutrino events, respectively.  
}
\label{fig:03}       
\vspace{-1cm}
\rotatebox{90}{%
\resizebox{0.6\textwidth}{!}{%
  \includegraphics{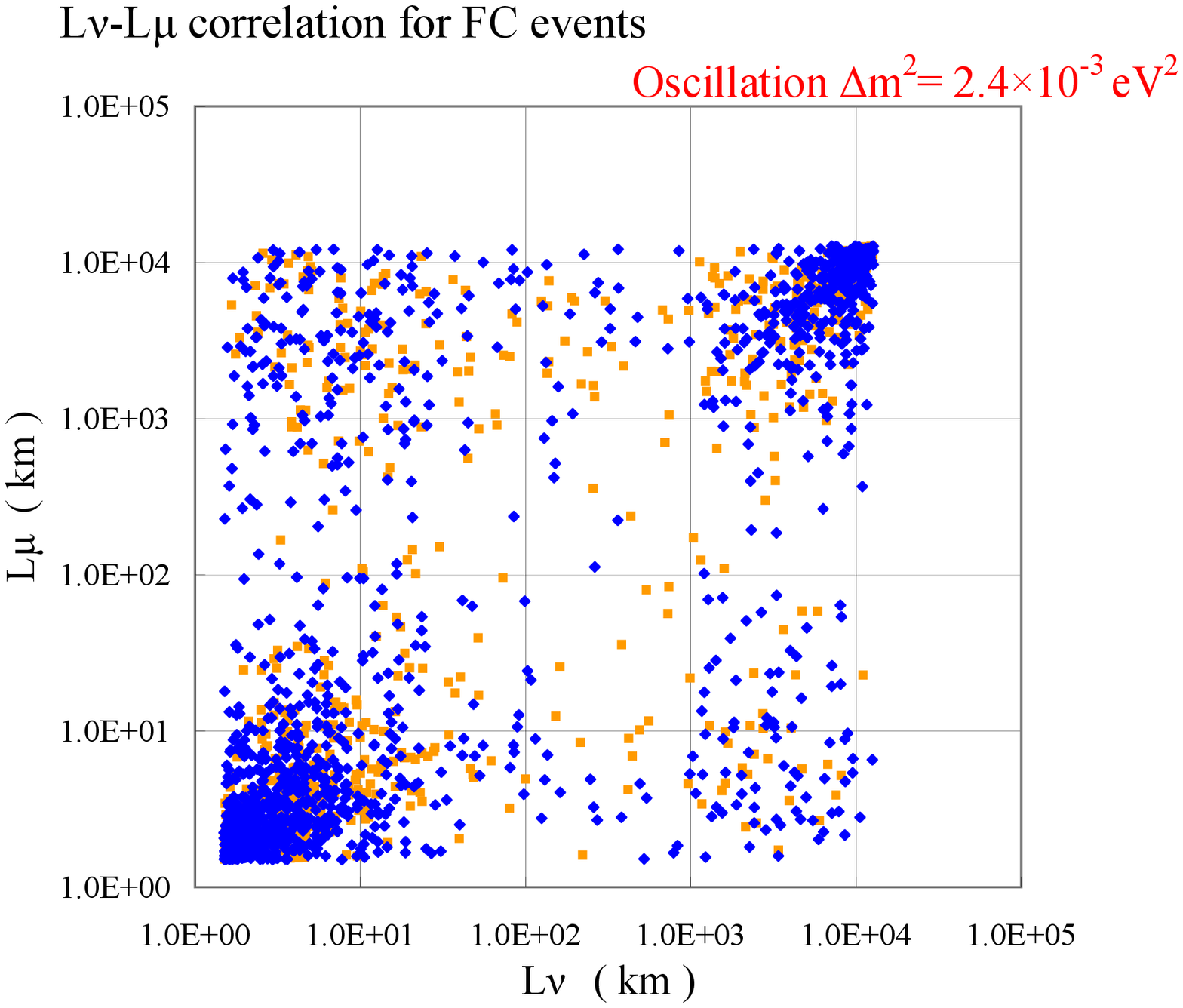}
}}
\vspace{-1.2cm}
\caption{Correlation diagram for $L_{\nu}$ and $L_{\mu}$ with 
the oscillation for 1489.2 live days.
The blue points and orange points denote neutrino events and 
ani-neutrino events, respectively.  
}
\label{fig:04}
\end{center}
\end{figure}

In Figure~2, we give the procedure for our numerical computer experiment.
In our numerical computer experiment, we obtain 
$L_{\nu}$, $L_{\mu}$, $E_{\nu}$ and $E_{\mu}$.
Therefore, we could choose four combinations among them, namely
$L_{\nu}/E_{\nu}$, $L_{\mu}/E_{\mu}$, $L_{\mu}/E_{\nu}$
and $L_{\nu}/E_{\mu}$
for examination of maximum oscillations due to the neutrino oscillation. 
However, only the combination of  $L_{\mu}/E_{\mu}$ out of these four
combinations could be physically measurable.
In the present paper(I) and a subsequent paper(II), we
 examine 
the relation between each combination and the oscillatory
 signature which is directly connected with the maximum oscillation 
in the presence of the neutrino oscillation. 
Our numerical computer experiments are being carried out for 
two different live days; namely,  one is 1489.2 live days which is 
equal to the actual live days by Super-Kamiokande Experiment\cite{Ashie2} 
and the other is 14892 days, 
ten times as much as the Super-Kamiokande Experiment actual live days.
\section{Correlations between $L_{\nu}$ and $L_{\mu}$, and between $E_{\nu}$ and $E_{\mu}$}
\subsection{The correlation between $L_{\nu}$ and $L_{\mu}$}

Super-Kamiokande Collaboration 
adopt the approximation of
 $L_{\nu} \approx L_{\mu}$ in their $L/E$ analysis
\cite{Ashie1}.
 As described in the preceding paper\cite{part1},
 \noindent
the relation between direction cosine of the incident neutrino, 
$(\ell_{\nu(\bar{\nu})}, m_{\nu(\bar{\nu})}, n_{\nu(\bar{\nu})} )$, and 
that of the corresponding emitted lepton, $(\ell_{\rm r}, m_{\rm r}, 
n_{\rm r})$, 
for a given scattering angle, $\theta_{\rm s}$,  
and its azimuthal angle, $\phi$, resulting from QEL is given as \\

\vspace*{-0.5cm}
\hspace*{-0.5cm}
\begin{equation}
\left(
         \begin{array}{c}
             \ell_{\rm_{\mu}} \\
             m_{\rm_{\mu}} \\
             n_{\rm_{\mu}}
         \end{array}
       \right)
           =
       \left(
         \begin{array}{ccc}
            \displaystyle \frac{\ell n}{\sqrt{\ell^2+m^2}} & 
            -\displaystyle 
            \frac{m}{\sqrt{\ell^2+m^2}}        & \ell_{\nu(\bar{\nu})} \\
            \displaystyle \frac{mn}{\sqrt{\ell^2+m^2}} & \displaystyle 
            \frac{\ell}{\sqrt{\ell^2+m^2}}     & m_{\nu(\bar{\nu})}    \\
                        -\sqrt{\ell^2+m^2} & 0 & n_{\nu(\bar{\nu})}
         \end{array}
       \right)
       \left(
          \begin{array}{c}
            {\rm sin}\theta_{\rm s}{\rm cos}\phi \\
            {\rm sin}\theta_{\rm s}{\rm sin}\phi \\
            {\rm cos}\theta_{\rm s},
          \end{array}
       \right)
\label{eqn:2}
\end{equation}
\noindent where $n_{\nu(\bar{\nu})}={\rm cos}\theta_{\nu(\bar{\nu})}$, and 
$n_{\rm_{\mu}}={\rm cos}\theta_{\mu}$. 
Here, $\theta_{\mu}$ is the zenith angle of the emitted muon.
\footnote
{The correlation between $cos\theta_{\nu}$ and $cos\theta_{\mu}$ is 
given in the Figure~12 in the preceding paper\cite{part1}.} 

$L_{\nu}$ and $L_{\mu}$ are functions of the direction cosine of the 
incident neutrino, $cos\theta_{\nu}$, and that of emitted muon,
 $cos\theta_{\mu}$, respectively and they are given as,
$$
L_{\nu}= R_g \times ( r_{SK} cos\theta_{\nu} +
\sqrt{ r_{SK}^2 cos^2\theta_{\nu} + 1 - r_{SK}^2} )  \,\,\,\,(3-1)
$$
$$
L_{\mu}= R_g \times ( r_{SK} cos\theta_{\mu} +
\sqrt{ r_{SK}^2 cos^2\theta_{\mu} + 1 - r_{SK}^2} )  \,\,\,\,(3-2)
$$
\noindent where $R_g$ is the radius of the Earth and 
$r_{SK}=1-D_{SK}/R_g$, with the depth, $D_{SK}$, 
of the Super-Kamiokande
Experiment detector from the surface of the Earth.
It should be noticed that the $L_{\nu}$ and $L_{\mu}$ are regulated by
both the 
energy spectrum of the incident neutrino and the production spectrum of 
the muon and, consequently, their mutual relation is influenced by either 
the absence of the oscillation or the presence of the oscillation which 
depend on the combination of the oscillation parameters.

  \begin{figure}
\begin{center}
\vspace{-1cm}
\rotatebox{90}{%
\resizebox{0.55\textwidth}{!}{%
  \includegraphics{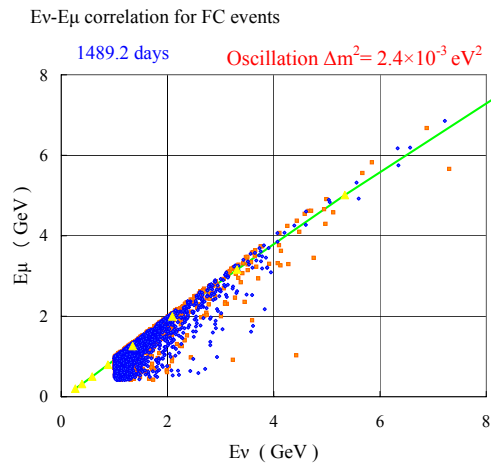}
}}
\vspace{-1.7cm}
\caption{The correlation diagram 
between $E_{\nu}$ and $E_{\mu}$ for the
oscillation.  The continous line denotes the polynomial
expression by the Super-Kamiokande
Collabolation.
}
\label{fig:06}       
\end{center}
\end{figure}

 In Figure~3, we give the correlation diagram between
$L_{\nu}$ and $L_{\mu}$ among {\it Fully Contained Events}
 for the 1489.2 live days in the absence of neutrino oscillation
which corresponds to the actual Super-Kamiokande Experiment\cite{Ashie2}.  
The aggregate of the (anti-) neutrino events which correspond to a definite 
combination of $L_{\nu}$ and $L_{\mu}$ are essentially classified into 
four groups in the following:

The group A is defined as the aggregate for neutrino events in which 
both $L_{\nu}$ and $L_{\mu}$ are rather smaller. 
It denotes that the downward neutrinos produce the downward muons.
 In this case, the energies of the produced muons are near the 
energies of the incident neutrinos due to smaller scattering angles.

 The group B is defined as the aggregate for neutrino events in which 
both $L_{\nu}$ and $L_{\mu}$ are rather larger.
It denotes that the upward neutrinos  produce upward muons.
In this case, the energy relation between the incident neutrinos and 
the produced muons is essentially the same as in the group A, because the 
flux of the upward neutrino events is symmetrical to that of the 
downward neutrino events in the absence of the neutrino oscillation.

The group C is defined as the aggregate for neutrino events in 
which $L_{\nu}$ are rather smaller and $L_{\mu}$ are rather larger.
 It denotes that the downward neutrinos produce the 
upward muons, namely by the possible effect reusulting from
 both backscattering and azimuthal angle in QEL,
 which is considered in the preceding paper
\cite{part1}.
 In this case, the energies of the produced muon are rather smaller 
than those of the energies of the incident neutrinos 
due to larger scattering angles.

The group D is defined as the aggregate for the neutrino events in 
which $L_{\nu}$ are rather larger and $L_{\mu}$ are rather smaller.
 It denotes that the upward the neutrinos produce the 
downward muons The energy relation between the incident neutrinos and 
the produced muon is essentially the same as in the group C in the absence 
of the neutrino oscillation.

It is clear from the Figure~3 that there exist the symmetries  between 
Group A and Group B, and also between Group C and Group D, 
 which reflect  
the symmetry between the upward neutrino flux and the downward neutrino 
one for null oscillation.  

 In Figure~4, we give the correlation between $L_{\nu}$ and $L_{\mu}$
under their neutrino oscillation parameters, say,
$\Delta m^2 = 2.4\times 10^{-3}\rm{ eV^2}$ and $sin^2 2\theta=1.0$
\cite{Ashie2}. 
In the presence of the neutrino oscillation, the property of the 
symmetry which holds in the absence of the neutrino oscillation 
(see [group A and group B] and/or [group C and group D] in Figure~3)
is lost due to the different incident neutrino fluxes in the upward 
direction and downward one. 
If we compare group A with group B, the event 
number in the group B (upward $\nu$ $\rightarrow$ upward $\mu$)
is smaller than that in the group A
(downward $\nu$ $\rightarrow$ downward $\mu$),
 which comes from smaller flux of the upward neutrinos. 
The similar relation between the group C 
(downward $\nu$ $\rightarrow$  upward $\mu$ ) and the group D 
(upward $\nu$ $\rightarrow$  downward $\mu$) is held in Figure~4.

 Summarizing the characteristics among the groups A to D in the 
Figures~3 and 4, we could conclude that [group A and group B] and 
[group C and group D] are in symmetrical situations in the absence 
of the neutrino oscillation,
while such the symmetrical situation is lost in the presence 
of the neutrino oscillation.
  Also, it is clear from the Figures~3 
and 4 that $L_{\nu}\approx L_{\mu}$,  
{\it the SK assumption on the direction} does not hold 
both in the absence of the neutrino oscillation and 
in the presence of the neutrino oscillation
even if statistically.

Here, it should be noticed that the approximation of
$L_{\nu}\approx L_{\mu}$ could not hold completely 
in the region C and region D.
The event number in the group C and group D could not be neglected 
among the total event number concerned. 
 In these regions, neutrino events consist of those with 
backscattering and/or neutrino events 
in which the neutrino directions
are horizontally 
downward (upward) but their produced muon turn to be upward (downward) 
resulting from 
the effect of azimuthal angles in QEL, respectively. 

\subsection{The correlation between $E_{\nu}$ and $E_{\mu}$}
 Super-Kamiokande Collaboration estimate $E_{\nu}$ from $E_{\mu}$,
 the visible energy of the muon,
 from their Monte Carlo simulation,
by the following equation\cite{Ishitsuka}(p.135) : 

$$
E_{\nu,SK}= E_{\mu}\times(a + b\times x + c\times x^2 + 
d\times x^3 ),  \,\,\,(4)
$$

\noindent where $x = log_{10}(E_{\mu})$.

 The idea that $E_{\nu}$ could be approximated as the polynomial means 
that there is unique relation between $E_{\nu}$ and $E_{\mu}$.
 However, in the light of stochastic characters inherent in the 
incident neutrino energy spectrum and the production spectrum of 
the muon, such the treatment is not suitable theoretically,
 which may kill real correlation effect between the incident 
neutrino energy and the emitted muon energy.
 In Figure~5, we give the correlation between $E_{\nu}$ and $E_{\mu}$ 
together with that obtained from the polynomial expression by 
the Super-Kamiokande Collaboration under 
the their neutrino oscillation parameters. 
It is clear from the figure that the part of the lower energy 
incident neutrino deviates largely from the approximated formula,
 which reflects explicitly the stochastic character of the QEL.
We could give the similar relation for null oscillation, the shape of 
which may be different from that with the oscillation due to the 
difference in the incident neutrino energy spectrum.

\begin{figure}
\begin{center}
\rotatebox{90}{%
\resizebox{0.5\textwidth}{!}{%
  \includegraphics{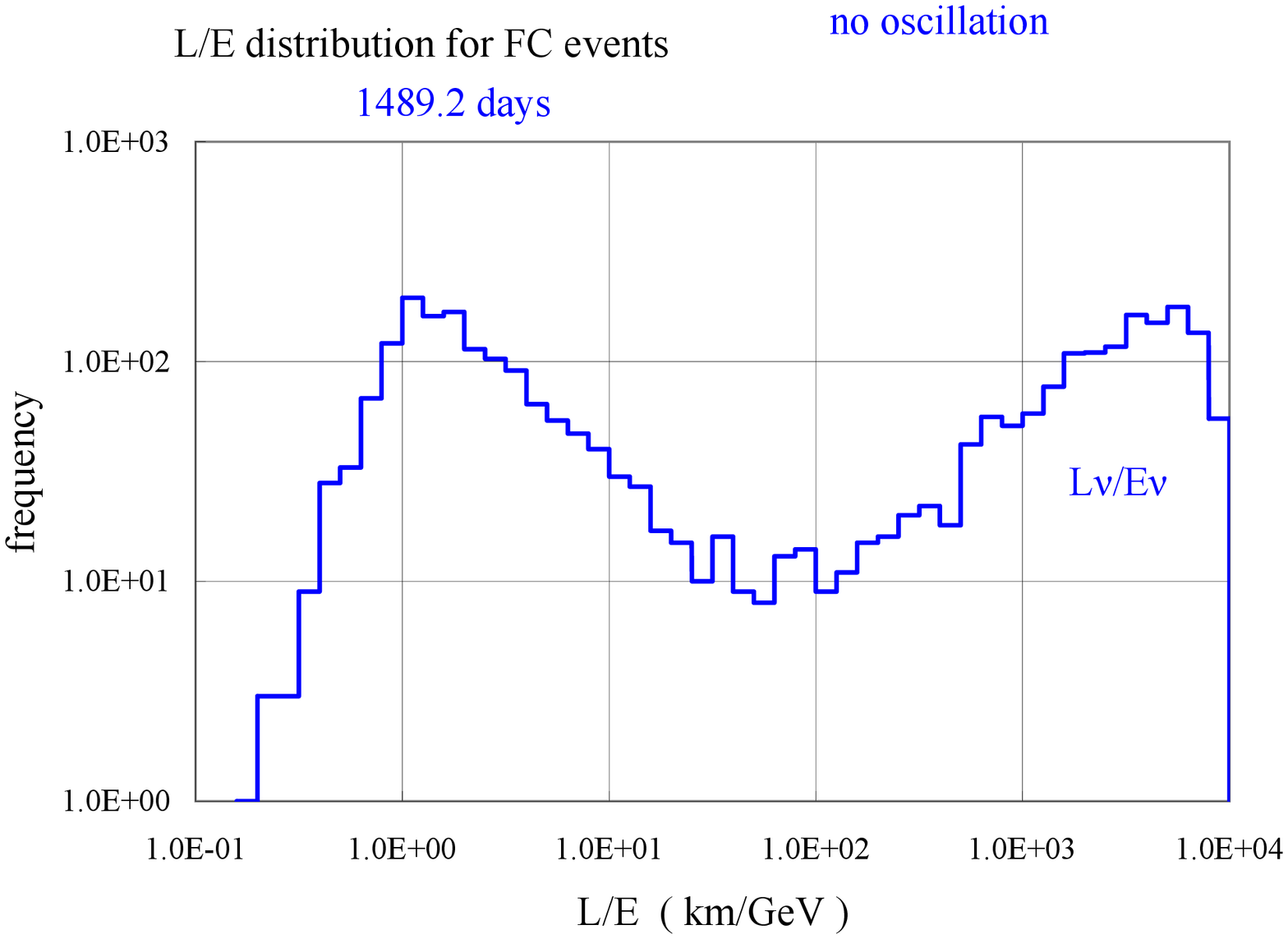}
}}
\caption{$L_{\nu}/E_{\nu}$ distribution without oscillation
for 1489.2 live days.}
\label{fig:07}       
\rotatebox{90}{%
\resizebox{0.5\textwidth}{!}{%
  \includegraphics{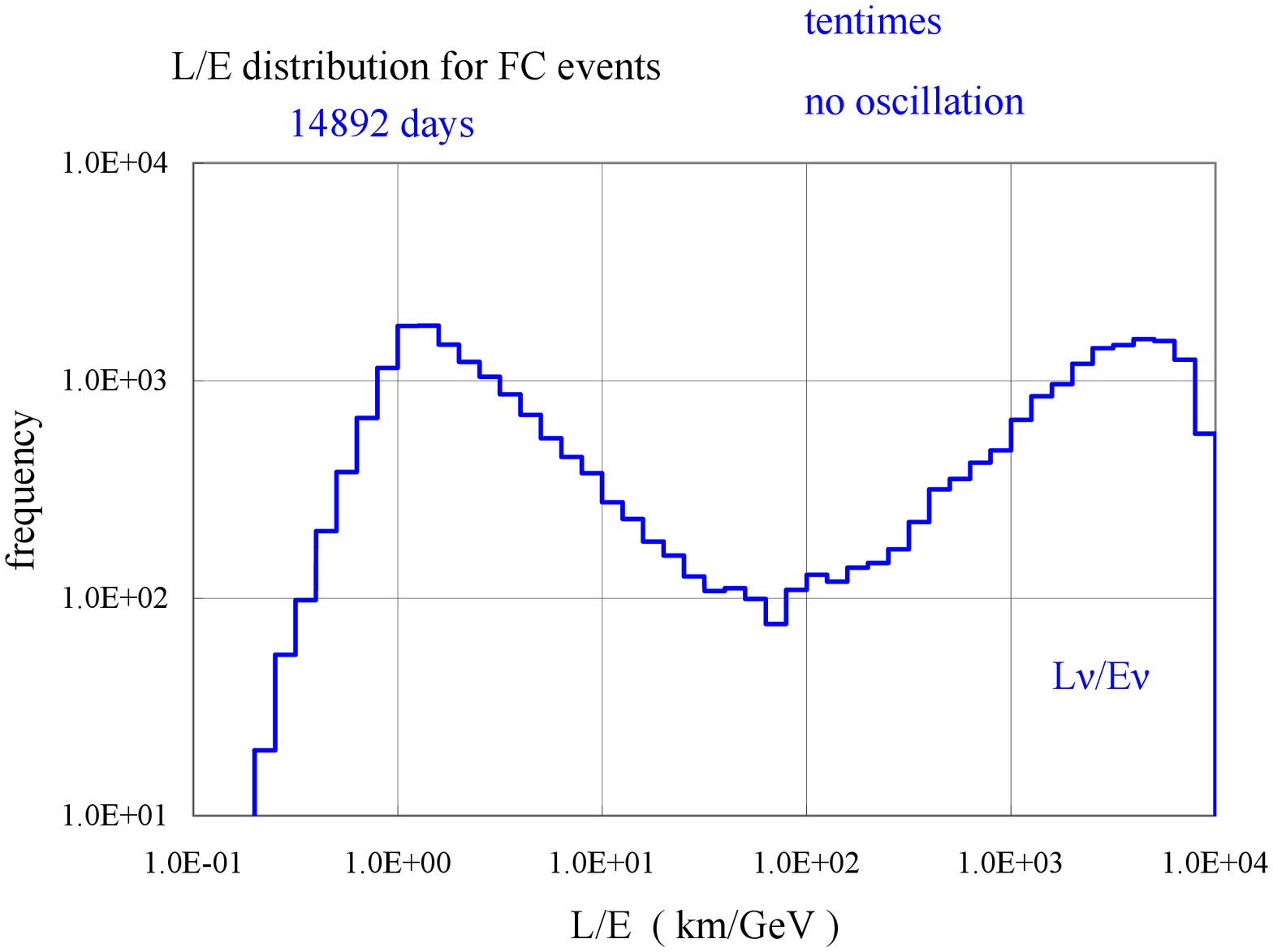}
}}
\caption{$L_{\nu}/E_{\nu}$ distribution without oscillation
for 14892 live days.}
\label{fig:08}       
\end{center}
\end{figure}

\begin{figure}
\begin{center}
\vspace{-0.5cm}
\rotatebox{90}{%
\resizebox{0.5\textwidth}{!}{%
  \includegraphics{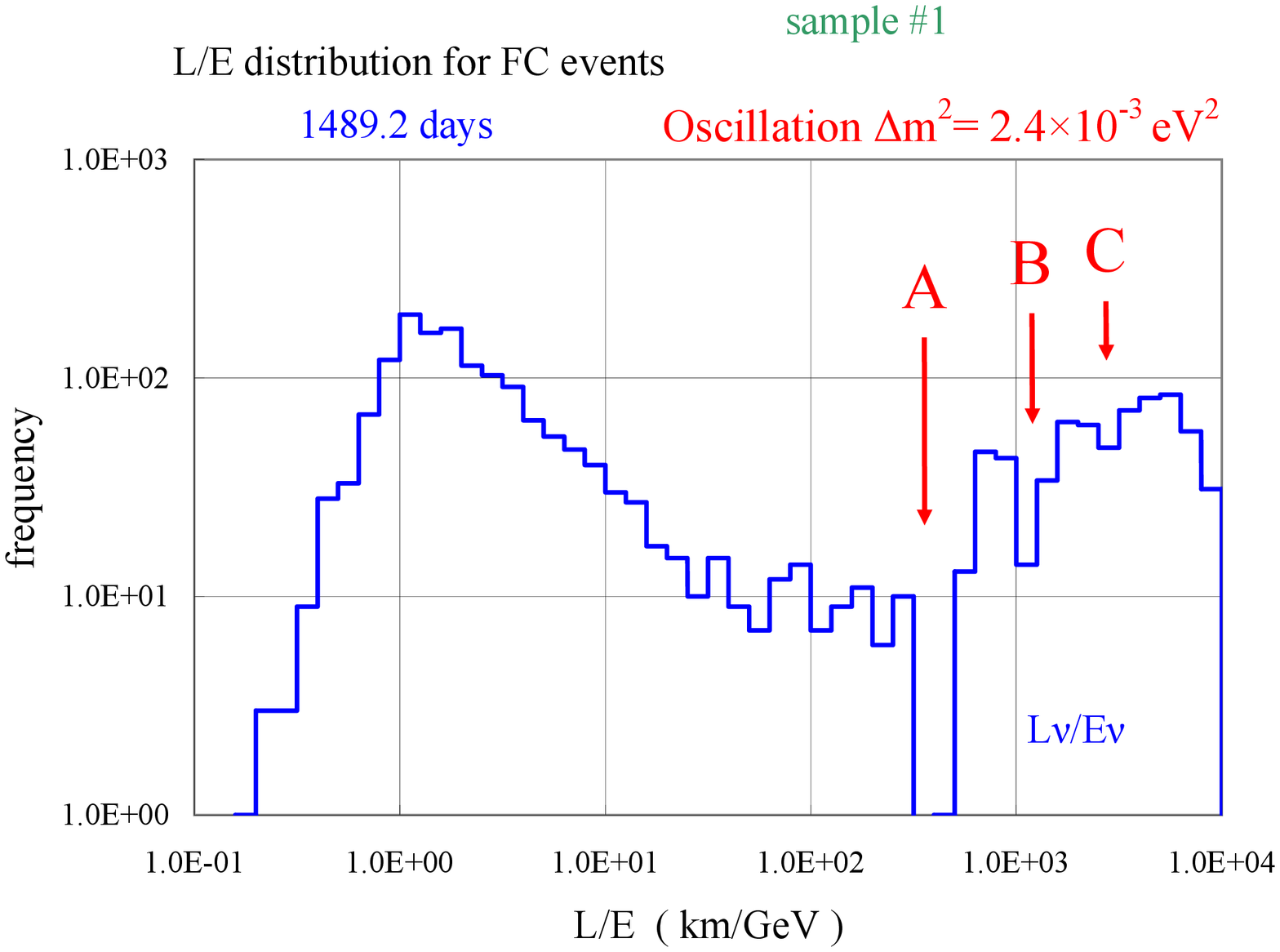}
}}
\vspace{-0.5cm}
\caption{$L_{\nu}/E_{\nu}$ distribution with the oscillation
for 1489.2 live days, sample~1.}
\label{fig:09}       
\vspace{-0.5cm}
\rotatebox{90}{%
\resizebox{0.5\textwidth}{!}{%
  \includegraphics{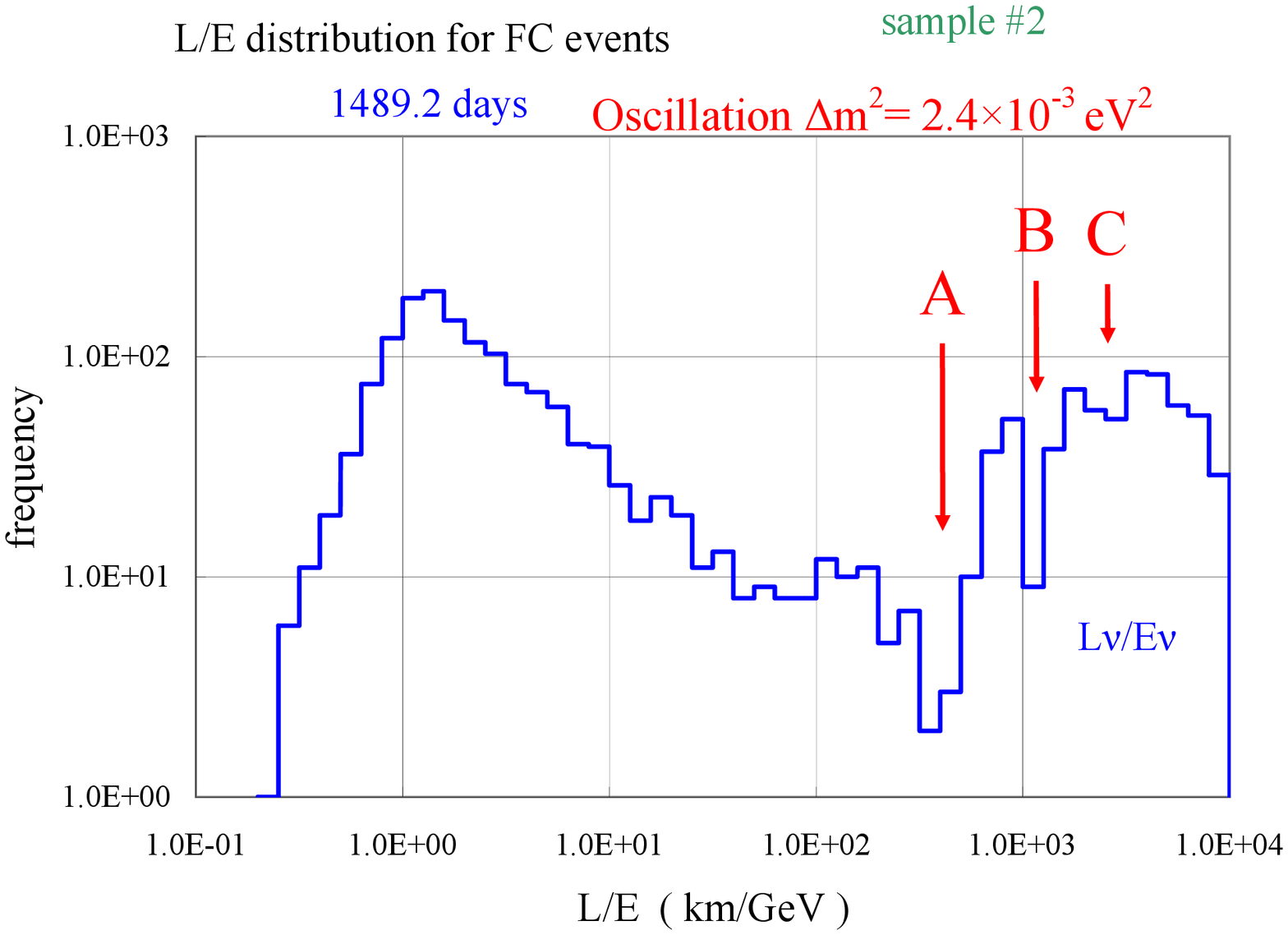}
}}
\vspace{-0.5cm}
\caption{$L_{\nu}/E_{\nu}$ distribution with oscillation
for 1489.2 live days, sample~2.}
\label{fig:091}       
\vspace{-0.5cm}
\rotatebox{90}{%
\resizebox{0.5\textwidth}{!}{%
  \includegraphics{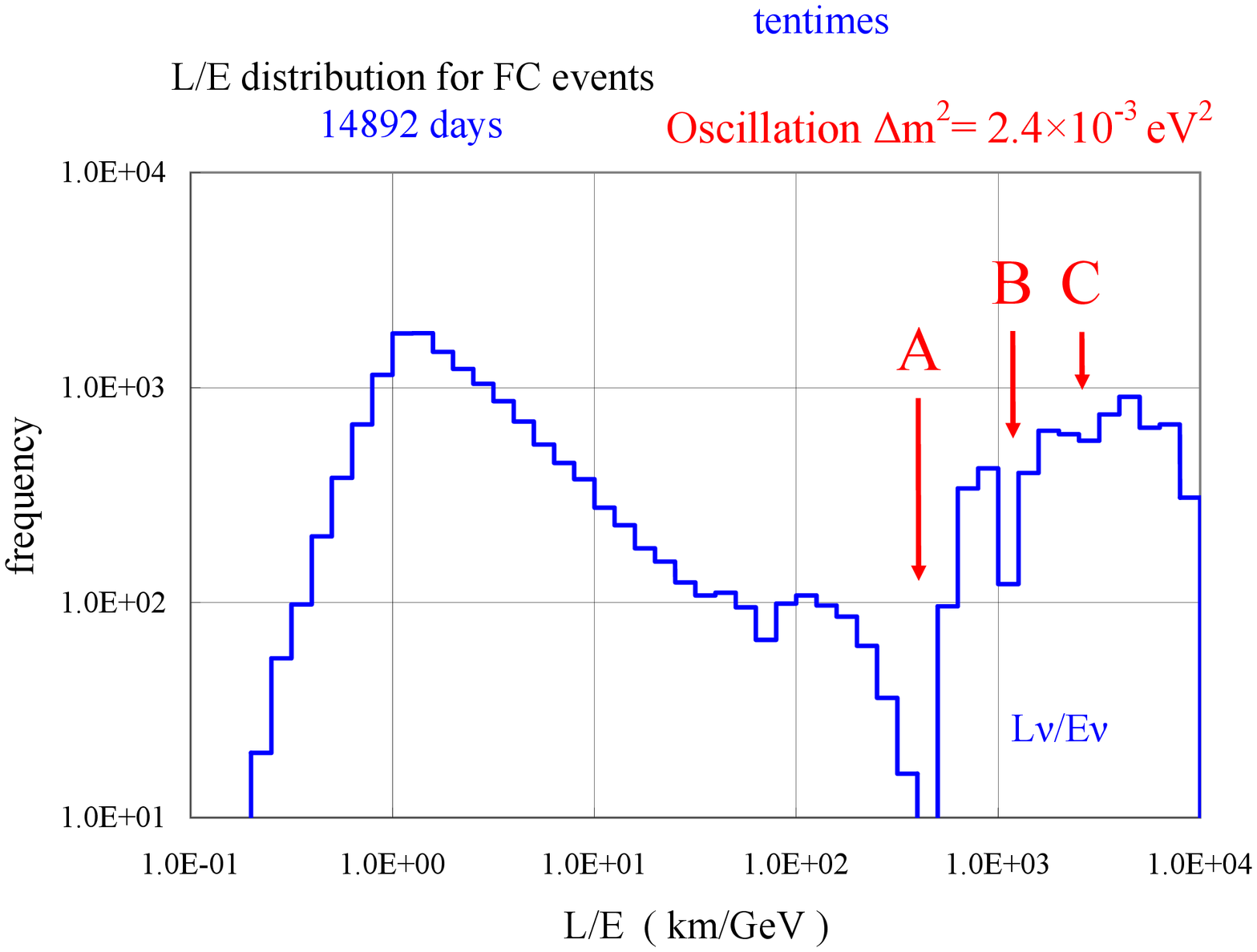}
}}
\vspace{-0.5cm}
\caption{$L_{\nu}/E_{\nu}$ distribution with oscillation
for 14892 live days.}
\label{fig:10}       
\end{center}
\end{figure}

\section{$L_{\nu}/E_{\nu}$ Distributions in Our Numerical Experiment}
\subsection{For null oscillation}
In Figures~6 and 7, we give $L_{\nu}/E_{\nu}$ distribution without 
oscillation for 1489.2 live days (SK live days\cite{Ashie2}) and 14892 
live days, respectively. Both distributions show the sinusoidal-like
 character for $L_{\nu}/E_{\nu}$ distribution, namely, 
the appearance of the top and the dip, appear even for null
 oscillation. The uneven histograms in Figure~6, comparing 
 with those in Figure~7, show that the statistics of 
the Figure~6 is not enough compared with that of the Figure~7.
Roughly speaking, smaller $L_{\nu}/E_{\nu}$ 
correspond to the contribution from downward neutrinos, larger one 
correspond to that from upward neutrino and $L_{\nu}/E_{\nu}$ near the dip 
correspond to the horizontal neutrinos, although the real situation is
more complicated, because the backscattering effect 
as well as the azimuthal angle effect could not be neglected.
From Figure~7, we understand that the dip around 70~km/GeV
denotes the contribution from the horizontal direction.
\begin{figure}
\begin{center}
\vspace{-1cm}
\rotatebox{90}{%
\resizebox{0.6\textwidth}{!}{%
  \includegraphics{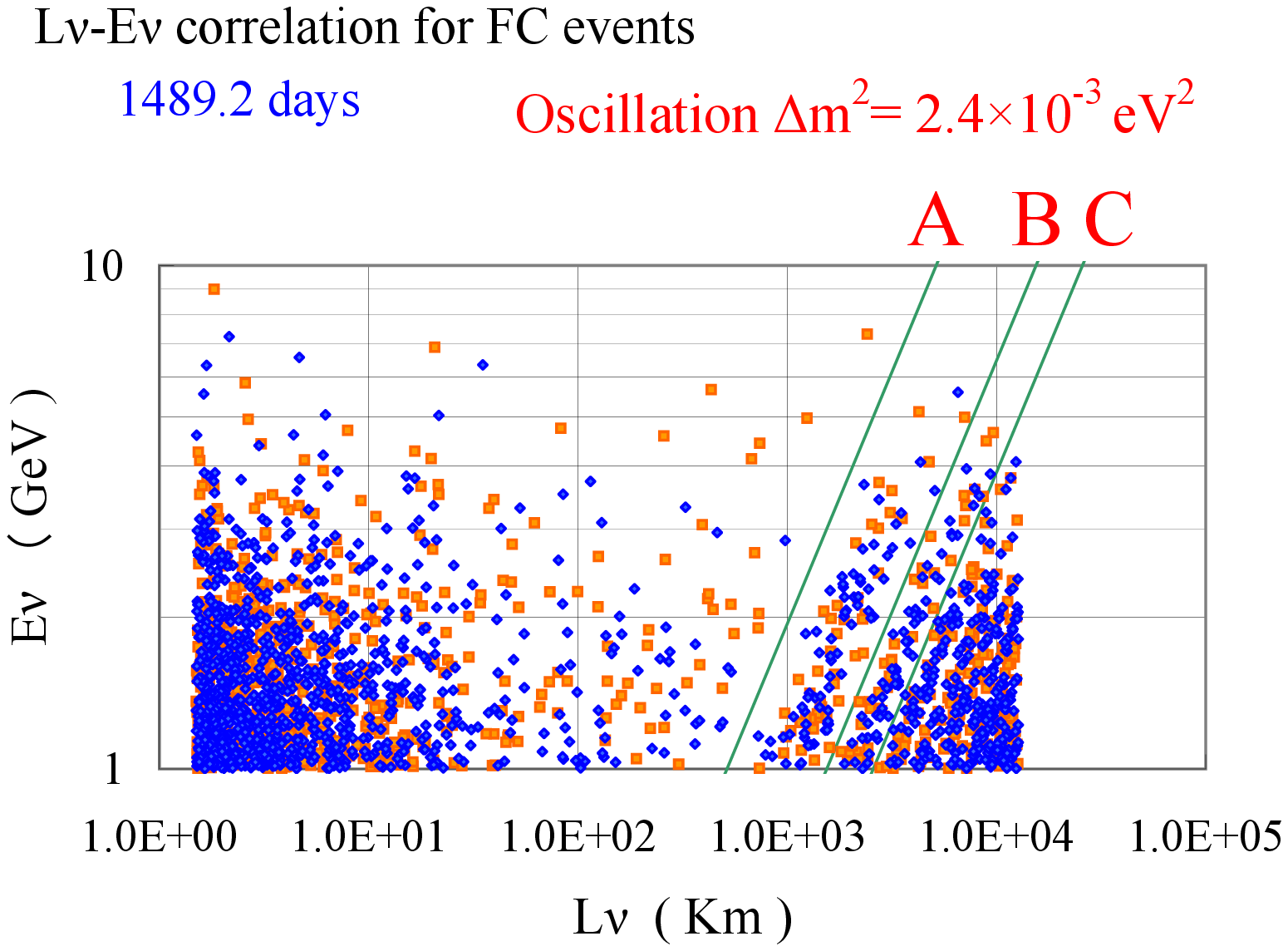}
}}
\vspace{-3cm}
\caption{The correlation diagram between $L_{\nu}$ and $E_{\nu}$
with oscillation for 1489.2 live days.}
\label{fig:11}       

\vspace{-1cm}
\rotatebox{90}{%
\resizebox{0.6\textwidth}{!}{%
  \includegraphics{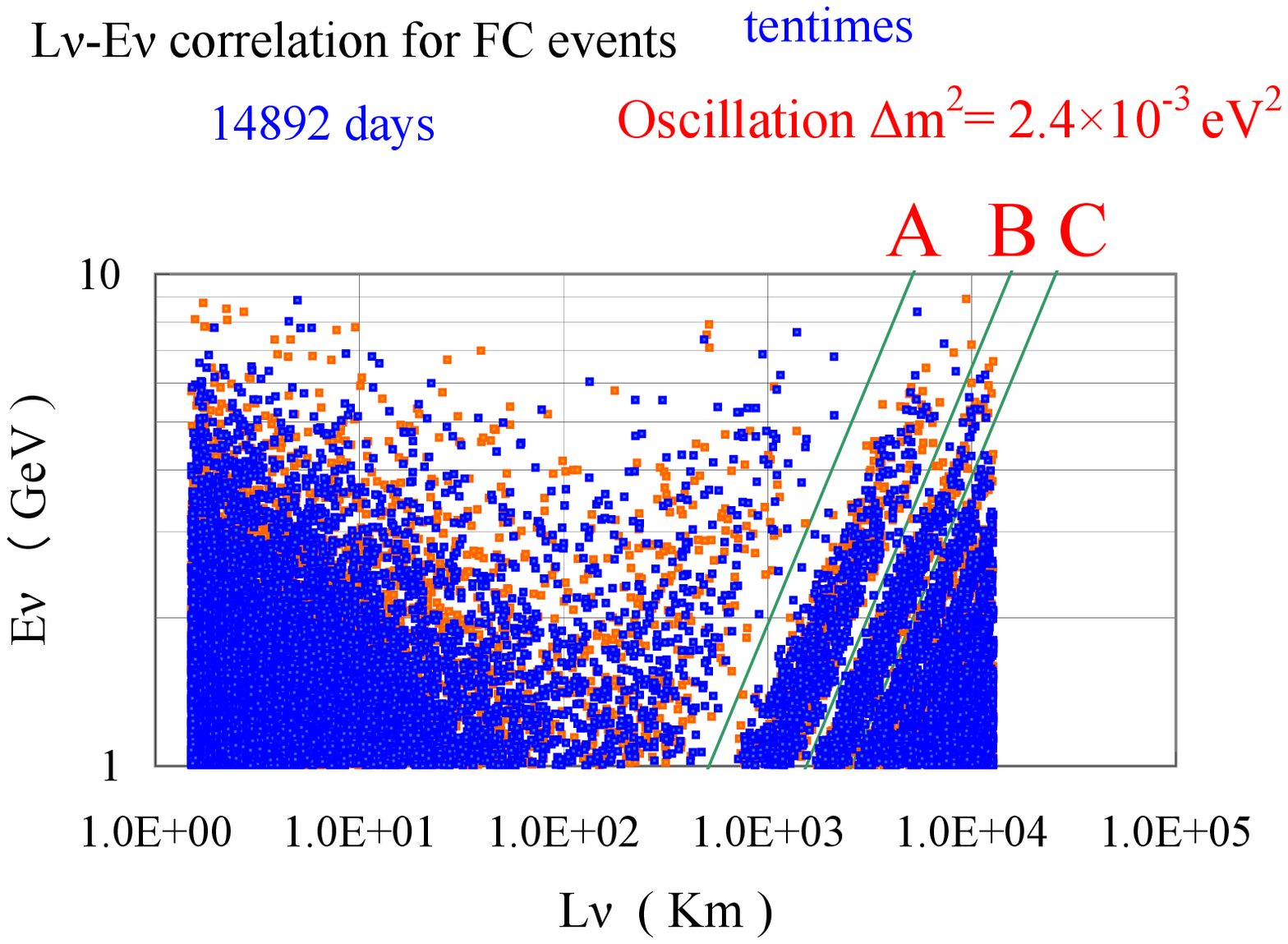}
}}
\vspace{-3cm}
\caption{The correlation diagram between $L_{\nu}$ and $E_{\nu}$
with oscillation for 14892 live days.}
\label{fig:12}       
\end{center}
\end{figure}
\begin{figure}
\begin{center}
\rotatebox{90}{%
\resizebox{0.5\textwidth}{!}{%
  \includegraphics{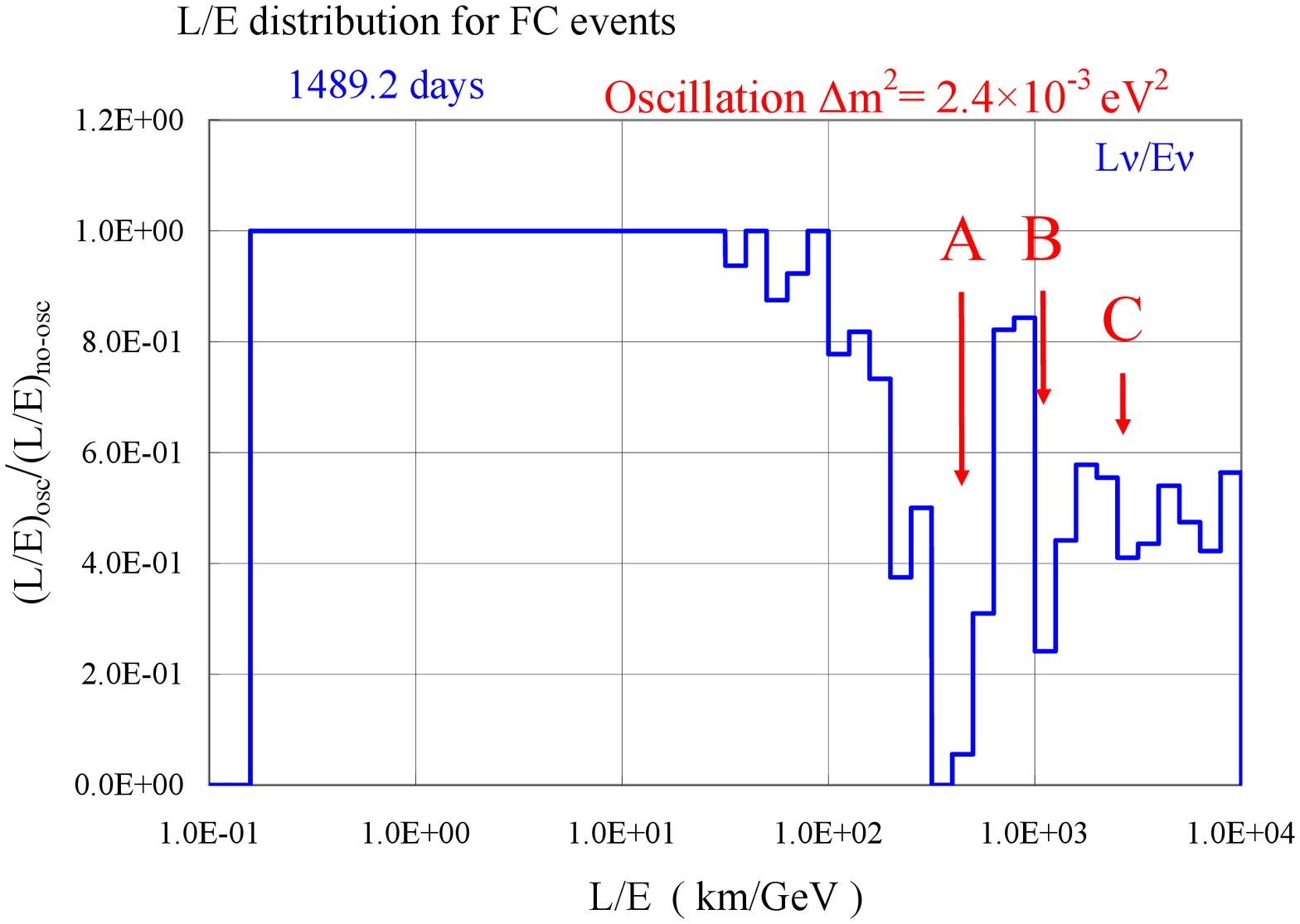}
}}
\caption{The ratio of 
$(L_{\nu}/E_{\nu})_{osc}/(L_{\nu}/E_{\nu})_{null}$
 for 1489.2 live days.}
\label{fig:115}       


\rotatebox{90}{%
\resizebox{0.5\textwidth}{!}{%
  \includegraphics{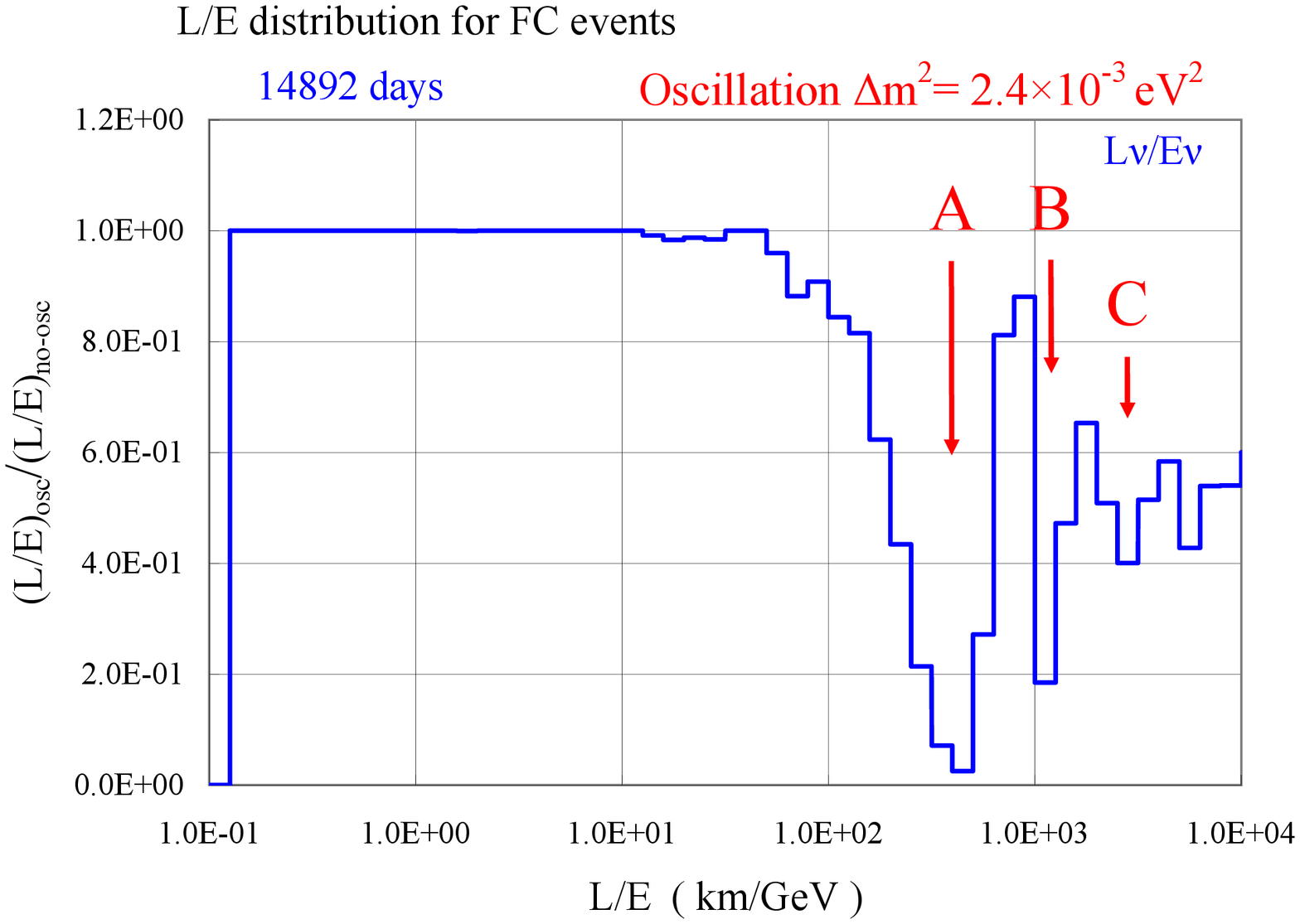}
}}
\caption{The ratio of 
$(L_{\nu}/E_{\nu})_{osc}/(L_{\nu}/E_{\nu})_{null}$
 for 14892 live days.}
\label{fig:125}       
\end{center}
\end{figure}

\subsection{For the oscillation (SK oscillation parameters)}
Super-Kamiokande Collaboration analyze experimental data consisting of 
{\it Fully Contained Events} and {\it Partially 
Contained Events} (single ring mu-like events and multi-ring 
mu-like events) on the $L/E$ distribution, expecting to find their 
oscillatory signature,
say, the presence of maximum osccillations,
 and they claim that they 
find a dip near 500 km/GeV which is consistent with their 
neutrino oscillation parameters 
$\Delta m^2 = 2.4\times 10^{-3}\rm{eV^2}$ and 
$sin^2 2\theta=1.0$\cite{Ashie2}. 

The survival probability of a given flavor, such as $\nu_{\mu}$,
 is given by
$$
P(\nu_{\mu} \rightarrow \nu_{\mu}) 
= 1- sin^2 2\theta sin^2
(1.27\Delta m^2 L_{\nu} / E_{\nu} ).         \,\,\,(5)  
$$                                                    

\noindent Then, for maximum oscillation under SK neutrino oscillation 
parameters,  we have
$$
 1.27\Delta m^2 L_{\nu} / E_{\nu} = (2n+1)\times\frac{\pi}{2},  \,\,\,\,\,\,(6)  
$$                                                    
where $\Delta m^2 = 2.4\times 10^{-3}\rm{eV^2}$.
From the eqation, we have the following values of
$L_{\nu} / E_{\nu}$ for maximum oscillations.
\begin{eqnarray}
     L_{\nu}/E_{\nu} =
  515 {\rm km/GeV}\,\,\,   for\,\, n=1   \,\,\, (7-1)\nonumber\\
 1540 {\rm km/GeV}\,\,\,   for\,\, n=2   \,\,\, (7-2)\nonumber\\ 
 2575 {\rm km/GeV}\,\,\,   for\,\, n=3   \,\,\, (7-3) \nonumber\\
\nonumber
\end{eqnarray}

In Figures~8 and 9, we give the $L_{\nu} / E_{\nu}$ distributions with 
the oscillation for one unit of 1489.2 live days and the another unit 
of 1489.2 live days, respectively. 
In Figure~8, we could observe the first maximum oscillation 
(Eq.7-1, arrow A) and the second(Eq.7-2, arrow B) clearly, 
and the third (Eq.7-3, arrow C) faintly,
 which means not enough statistics for the confirming the existence 
of the third maximum oscillation.
 In Figure~9, we could observe similar situation for
 another 1489.2 live days as observed in Figure~8. 
 In Figure~10, we give the $L_{\nu} / E_{\nu}$ distribution with the  
oscillation for 14892 live days being ten times as much as those of
 Figures~8 and 9. 
In Figure~10, we could observe a more narrower dip which corresponds to 
the first maximum oscillation due to ten times statistics as much as
those of Figures~8 and 9. Also, we could observe the second 
maximum oscillation clearly, too. 
However, it is a little bit difficult to observe the 
third maximum oscillation for even larger statistics.

 In Figure~11, we give the correlation between $L_{\nu}$ and $E_{\nu}$
 which is obtained from the same data as in Figure~8. 
The linearly strip patterns in which there are absent from the 
combinations of $L_{\nu}$ and $E_{\nu}$, correspond to the first,
 the second and third maximum in the $L_{\nu} / E_{\nu}$ distribution.
  In Figure~12, we give the same correlation for 14892 days being ten 
times larger than the Super-Kamiokande Experiment actual live days.  
Compared Figure~12 with Figure~11, the larger statistics show more 
clear strip pattern which correspond to more clear existence of the
 maximum oscillation as in the Figure~10.
Lines A, B and C in Figures~11 and 12 
 correspond to Eqs.(7-1),(7-2) 
and (7-3) respectively. Surely, we observe dips characterized by these 
lines where we have no events. Namely, these figures show the presence of the maximum oscillation obtained from Eq.(6).\\

 From the point of view of the numerical computer experiment, 
the survival probability of a given flavor is give as the ratio of 
$(L_{\nu}/E_{\nu})_{osc}/(L_{\nu}/E_{\nu})_{null}$.
 In Figures~13 and Figure 14, these ratios are given for 1489.2 live 
days and 14892 live days being ten times as much as 1489.2 live days, 
respectively. These ratios indicate 
the presence of the maximum oscillations shown as arrows A, B and C.  
Comparing the ratio of 1489.2 live days with that of
14892 live days in the 
light of their smoothness, it seems that statistics with 1489.2 live 
days are not so beautiful compared to statistics with
 14892 live days.
 
Now, it is concluded from Figures 8 to 14 that our numerical 
computer experiments are carried out in a correct way, which shows 
really the sinusoidal flavor transition probability of neutrino 
oscillation under the neutrino oscillation parameters
by the Super-Kamiokande Collaboration, 
if they really exist.
 However, at the same time, it should be noticed that such neutrino 
oscillatory signature could not be observed really at all even if it
 really exists, because both the physical quantities,
$L_{\nu}$ and $E_{\nu}$, are not physical observables.

\section{Summary}
In our numerical computer experiment, we have found the existence 
of the maximum oscillation, 515km/GeV, 1540km/GeV and 2575km/GeV,
 under the neutrino oscillation parameter 
($\Delta m^2 = 2.4\times 10^{-3}\rm{ eV^2}$ and $sin^2 2\theta=1.0$
),  successfully, as they should be. 
It denotes that our numerical computer experiment have been done 
in correct way.
Both $L_{\nu}$ and $E_{\nu}$ could not be measurable,
 because neutrinos are neutral. 
Consequently, even if the neutrino oscillation exist really, 
we could not detect neutrino oscillation from 
$L_{\nu}/E_{\nu}$ analysis.
In the subsequent paper, we examine whether the maximum oscillation 
resulting from the neutrino oscillation through the analysis of 
other possible combinations of L/E, 
such as $L_{\nu}/E_{\mu}$, $L_{\mu}/E_{\nu}$ and
$L_{\mu}/E_{\mu}$,
can be detected or not.
\newpage
\newpage
\appendix
\section{\appendixname:{ }Detection of the Neutrino Events in the SK 
Detector and Their Interaction Points}
\setcounter{equation}{0}
\def\theequation{\Alph{section}\textperiodcentered\arabic{equation}}

\begin{figure}[b]
\begin{center}
\resizebox{0.45\textwidth}{!}{%
  \includegraphics{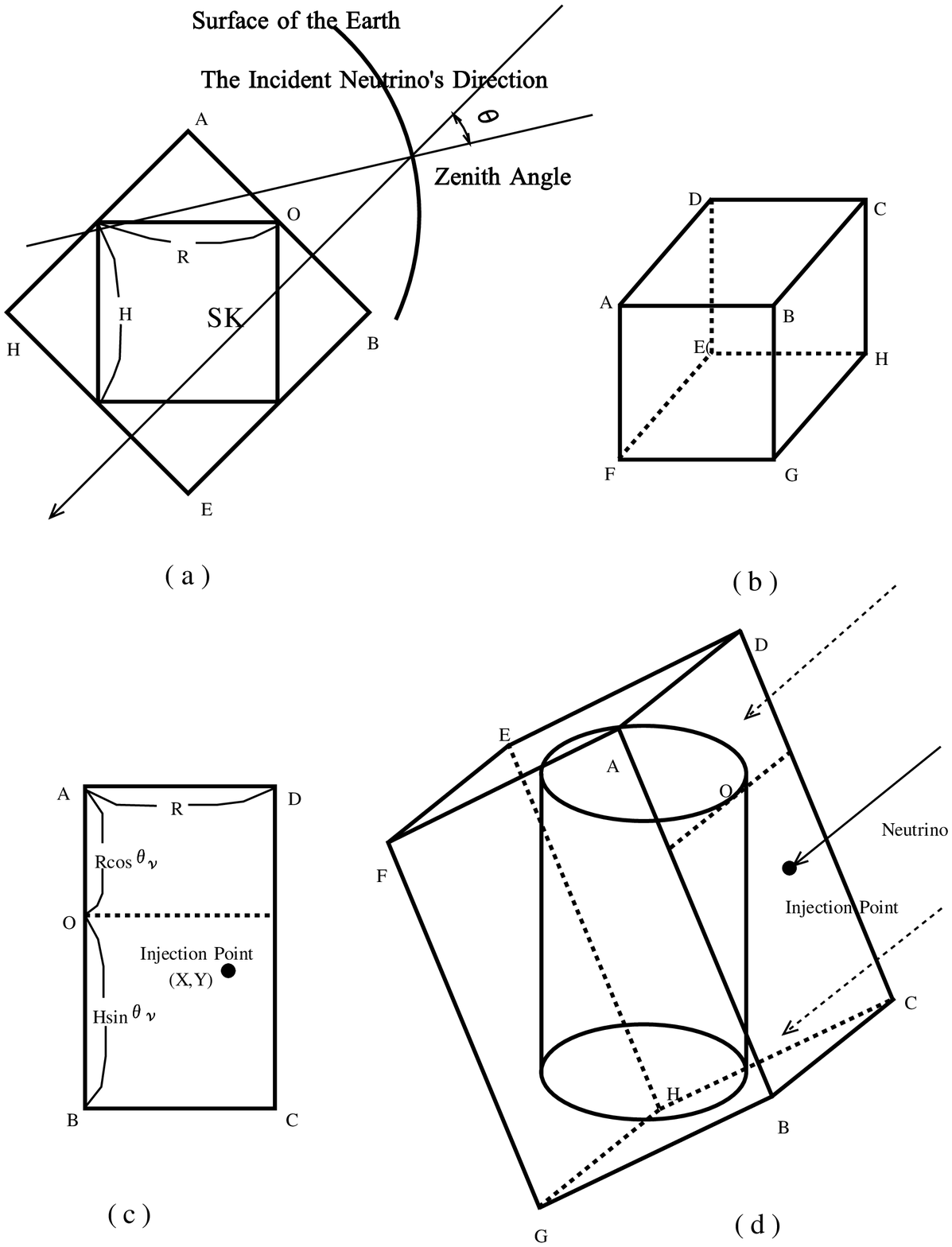}
  }
\end{center}
\caption{\label{fig:ABCD} Sampling procedure for neutrino evens injected into the detector}
\end{figure}

The plane ABCD is always directed vertically to the direction of the incident neutrino with a given zenith angle, which is shown in Fig. \ref{fig:ABCD}. The rectangular ABCDEFGH encloses the detector 
of the Super-Kamiokande Experiment
whose radius and height are denoted by \textit{R} and \textit{H}, respectively. The width and the height of the plane ABCD for a given zenith angle,  ${\theta_{\nu(\bar{\nu})}}$, are given as, \textit{R} and
\textit{R}$\cos\theta_{\nu(\bar{\nu})}$ + \textit{H}$\sin{\theta}_{\nu}$,
 respectively, which are shown in Fig. A1-(c).

Now, let us estimate the ratio of the number of the neutrino events 
inside the detector of the Super-Kamiokande Experiment
 to that in the rectangular ABCDEFGH. As the number of the neutrino 
events inside some material is proportional to the number of the nucleons in the material concerned. 
The number of the nucleons inside the SK detector $(\rho=1)$ is given as

     \begin{equation}
         N_{\rm sk}=\frac{\pi}{4}N_{\rm avoga}R^2H,
     \end{equation}
\noindent where $N_{\rm avoga}$ denotes the Avogadro number, and the number of the nucleons in the exterior of the detector inside ABCDEFGH is given as
\\

\begin{equation}
       N_{\rm extr}(\cos\theta_{\nu(\bar{\nu})}) =\rho N_{\rm avoga} 
       \biggl[ \biggl( 1-\frac{\pi}{4} \biggr)R^2H + 
        R(H^2 \!\! + \!\! R^2)\sin{\theta}_{\nu(\bar{\nu})}\cos\theta_{\nu(\bar{\nu})} \biggr],
\end{equation}

\noindent  where $\rho$ is the density of the rock which surrounds the SK detector. 

Then, the total number of the target in the rectangularABCDEFGH is given as
\\
    \begin{equation}
       N_{\rm tot}(\cos\theta_{\nu(\bar{\nu})}) =N_{\rm sk}+N_{\rm extr}(\cos\theta_{\nu(\bar{\nu})}) .
    \end{equation}

\noindent Here, we take 2.65 as ${\rho}$  (standard rock).\\

Then, $R_{\rm theor}$, the ratio of the number of the neutrino events in the SK detector to that in the rectangular ABCDEFGH is given as

   \begin{equation}
          R_{\rm theor}(\cos\theta_{\nu(\bar{\nu})})  = N_{\rm sk} /  N_{\rm tot}(\cos\theta_{\nu(\bar{\nu})})   .
   \end{equation}

\noindent We obtain $R_{\rm theor}$ for different values of $\cos{\theta_{\nu(\bar{\nu})}}$ given in the Table~1.
\\
  Here, we simulate neutrino events occured in the rectangular ABCDEFH, by using the atmospheric neutrino beam which falls down on the plane ABCD. Thus, $N_{\rm smaple}$, the sampling number of the (anti-)neutrino events inside the rectangular ABCDEFG for a given ${\cos\theta_{\nu(\bar{\nu})}}$ 
is given as
\\
   \begin{eqnarray}
\lefteqn{
       N_{\rm sample} (\cos\theta_{\nu(\bar{\nu})}) = N_{\rm tot}(\cos\theta_{\nu(\bar{\nu})})\times 
} \nonumber \\
&&      \int_{E_{\nu(\bar{\nu}),{\rm min}}}^{E_{\nu(\bar{\nu}),{\rm max}}}
       \sigma_{_{\ell(\bar{\ell})}}(E_{\nu(\bar{\nu})})N_{\rm int}(E_{\nu(\bar{\nu})},t,{\rm cos}\theta_{\nu(\bar{\nu})}) 
      {\rm d}E_{\nu(\bar{\nu})} \nonumber \\
      &&
   \end{eqnarray}

\noindent where $\sigma_{\ell(\bar{\ell})}(E_{\nu(\bar{\nu})})$  is the total cross section for (anti-)neutrino due to QEL, and 
$N_{\rm int}(E_{\nu(\bar{\nu})},t,{\rm cos}\theta_{\nu(\bar{\nu})}){\rm d}E_{\nu(\bar{\nu})}$ 
is the differential nutrino energy spectrum for the definite zenith angle, $\theta_{\nu(\bar{\nu})}$, in the plane ABCD. The injection points of the neutrinos in the plane ABCD are distributed over the plane randomly and uniformely and the injection points are determined from a pair of the uniform random  numbers between (0,1). They penetrate into the rectangular ABCDEFGH from the injection point in the plane ABCD and some of them may penetrate into the SK detector or may not, which depend on their injection point.

In the neutrino events which penetrate into the SK detector, their geometrical total track length, $T_{\rm track}$, are devided into three parts\\
    \begin{equation}
       T_{\rm track} = T_{\rm b} + T_{\rm sk} + T_{\rm a},
    \end{equation}
\noindent where $T_{\rm b}$ denotes the track length from the plane ABCD to the entrance point of the SK detector, $T_{\rm sk}$ denotes the track length inside the SK detector, and $T_{\rm a}$ denotes the track length from the escaping point of the SK detector to the exit point of the rectangular ABCDEF, 
and thus $T_{\rm track}$ denotes the geometrical length of the neutrino concerned in the rectangular ABCDEFGH. 
\newline

By the definition, the neutrinos concerned with $T_{\rm track}$ interact surely somewhere along the $T_{\rm track}$. Here, we are interested only in the interaction point which occurs along $T_{\rm sk}$.  We could determine the interaction point in the $T_{\rm sk}$  in the following.\\

We define the following quantities for the purpose.
   \begin{eqnarray}
         T_{\rm weight} &=& T_{\rm sk} + \rho(T_{\rm b} + T_{\rm a}),\\
         \rho_{\rm av}  &=& T_{\rm weight} / T_{\rm track},\\
         \xi_{\rho} &=& \rho_{\rm av} / \rho,\\                         
         \xi_{\rm sk}   &=& T_{\rm sk} / T_{\rm weight}.
    \end{eqnarray}

\begin{figure}
\begin{center}
\resizebox{0.55\textwidth}{!}{%
  \includegraphics{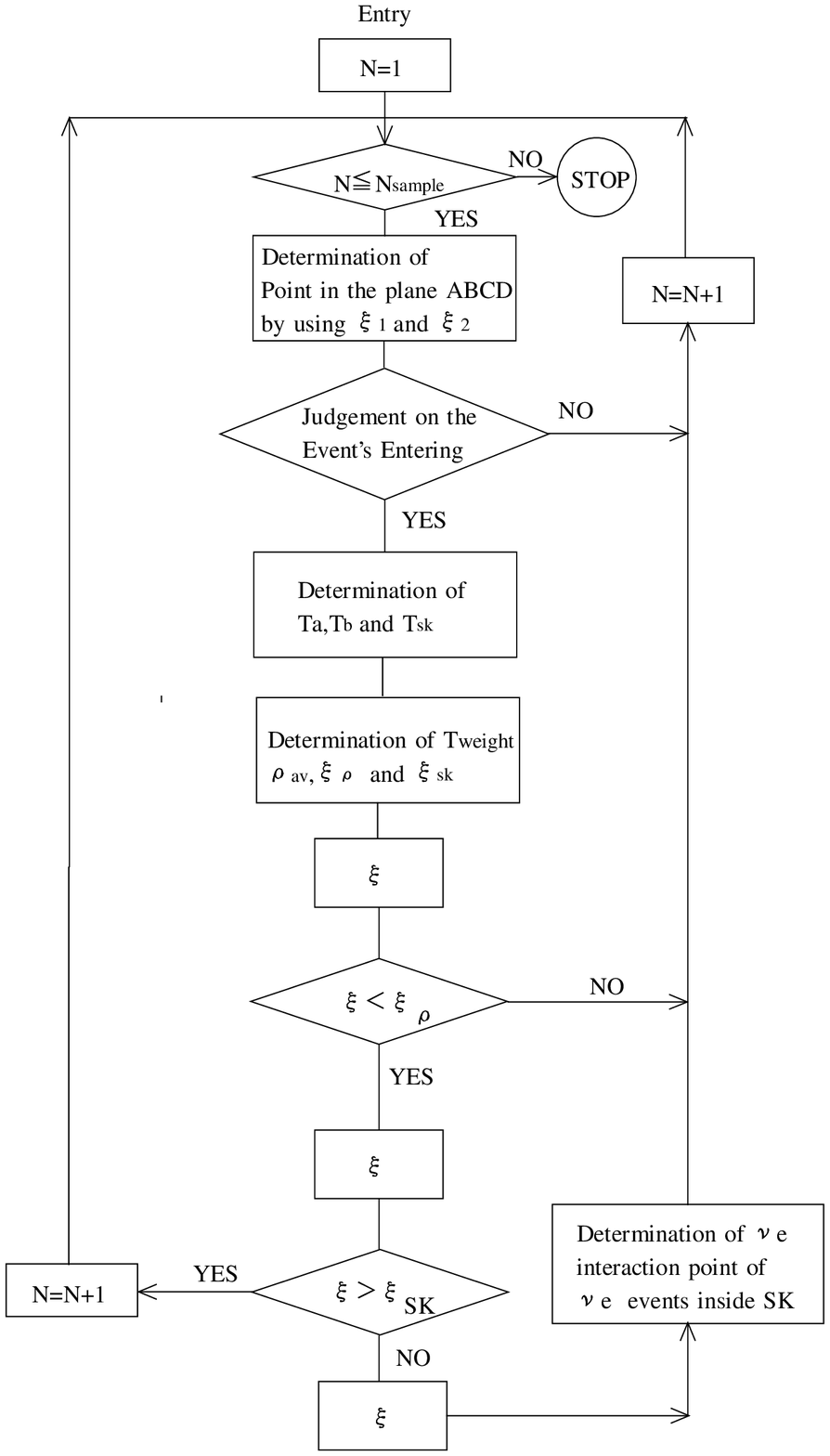}
  }
\end{center}
\vspace{-0mm}
\caption{\label{fig:16} Flow Chart for the determination of the interaction points of the neutrino events 
 inside the detector}
\end{figure}

The flow chart for the choice of the neutrino events in the SK detector and the determination of the interaction points inside the SK detector is given in Fig. \ref{fig:16}. Thus, we obtain neutrino events whose occurrence point is decided in the SK detector in the following.
\
     \begin{eqnarray}
       x_f &=& x_0,\\
       y_f &=& y_0 + \xi T_{\rm sk}\sin{\theta}_{\nu(\bar{\nu})}\\
       z_f &=& z_0 + \xi T_{\rm sk}\cos\theta_{\nu(\bar{\nu})}.
     \end{eqnarray}

If we carry out the Monte Carlo Simulation, following the flow chart in Fig. \ref{fig:16}, then, we obtain $N_{\rm event}$, the number of the neutrino events generasted in the SK detector. The ratio of the selected events to the total trial is given as
   \begin{equation}
          R_{\rm monte}(\cos\theta_{\nu(\bar{\nu})}) = N_{\rm event}(\cos\theta_{\nu(\bar{\nu})}) /  N_{\rm sample}( \cos\theta_{\nu(\bar{\nu})}).
   \end{equation}

\noindent Comparison between $R_{\rm theor}$ and $R_{\rm monte}$ in 
Table~A1 shows that our Monte Carlo procedure is valid. 
\begin{table}
\caption{{\label{tab:table3}}Occurrence probabilities of the neutrino events inside the SK detector for different $\cos \theta_{\nu}$'s. Comparison between $R_{\rm theor}$ and $R_{\rm monte}$. The sampling numbers for the Monte Carlo Simulation are, 1000, 10000, 100000, respectively.}
\begin{center}
\begin{tabular}{c c c c c}
\hline
cos $\theta_{\nu}$  & \, $R_{\rm {theor}}$ & \multicolumn{3}{c}{\, \, $R_{\rm {monte}}$} \\ \cline{3-5}
 & & \multicolumn{3}{c}{\, \, Sampling Number} \\ \cline{3-5}
 & & \, \, 1000 & 10000 & 100000 \\
\hline
0.000&\, 0.58002&\, \, 0.576&0.5750&0.57979\\
0.100&\, 0.41717&\, \, 0.425&0.4185&0.41742\\
0.200&\, 0.32792&\, \, 0.353&0.3252&0.32657\\
0.300&\, 0.27324&\, \, 0.282&0.2731&0.27163\\
0.400&\, 0.23778&\, \, 0.223&0.2329&0.23582\\
0.500&\, 0.21491&\, \, 0.206&0.2063&0.21203\\
0.600&\, 0.20117&\, \, 0.197&0.1946&0.19882\\
0.700&\, 0.19587&\, \, 0.193&0.1925&0.19428\\
0.800&\, 0.20117&\, \, 0.198&0.2002&0.20001\\
0.900&\, 0.22843&\, \, 0.230&0.2248&0.22803\\
1.00 &\, 0.58002&\, \, 0.557&0.5744&0.57936\\
\hline
\end{tabular}
\end{center}
\end{table}
\section*{References}


\end{document}